\documentclass[12pt]{iopart}

\usepackage{graphicx}

\begin{document}

\title{Noise induced oscillations in non-equilibrium steady state systems}{}
\author{I A Shuda$^1$, S S Borysov$^1$ and A I Olemskoi$^2$}
\address{$^1$Sumy State University, 2, Rimskii-Korsakov St., 40007
Sumy, Ukraine}

\address{$^2$Institute of Applied
Physics, Nat. Acad. Sci. of Ukraine, 58, Petropavlovskaya St.,
40030 Sumy, Ukraine}

\ead{alex@ufn.ru}

\begin{abstract}We consider effect of stochastic sources upon
self-organization process being initiated with creation of the limit cycle.
General expressions obtained are applied to the stochastic Lorenz system to
show that offset from equilibrium steady state can destroy the limit cycle at
certain relation between characteristic scales of temporal variation of
principle variables. Noise induced resonance related to the limit cycle is
found analytically to appear in non-equilibrium steady state system if the
fastest variations displays a principle variable, which is coupled with two
different degrees of freedom or more.
\end{abstract}

 \maketitle

\section{Introduction}\label{Sec.1}

Interplay between noise and non-linearity of dynamical systems \cite{1} is
known to arrive at crucial changing in behavior of systems displaying
noise-induced \cite{2,2a} and recurrence \cite{3a,3} phase transitions,
stochastic resonance \cite{4a,4}, noise induced pattern formation \cite{a,b},
noise induced transport \cite{c,2a} et cetera (see Ref. \cite{SSG}, for
review). The constructive role of noise on dynamical systems includes hopping
between multiple stable attractors \cite{e,f} and stabilization of the Lorenz
attractor near the threshold of its formation \cite{d,g}. Such type behavior is
inherent in finite systems which involve discrete entities (for instance, in
ecological systems individuals form population stochastically in accordance
with random births and deaths). Examples of substantial alteration of finite
systems under effect of intrinsic noises give epidemics \cite{11}--\cite{13},
predator-prey population dynamics \cite{5,6}, opinion dynamics \cite{10},
biochemical clocks \cite{15,16}, genetic networks \cite{14}, cyclic trapping
reactions \cite{9}, etc.

Within the phase-plane language, phase transitions po\-in\-ted out present the
simplest case, where a fixed point appears only. We are interested in studying
more complicated situation, when the system under consideration may display
oscillatory behavior related to the limit cycle appearing as a result of the
Hopf bifurcation \cite{17,18}. It has long been conjectured \cite{19} that in
some situations the influence of noise would be sufficient to produce cyclic
behavior \cite{20}. Recent consideration \cite{21} allows the relation between
the stochastic oscillations in the fixed point phase and the oscillations in
the limit cycle phase to be elucidated. Moreover, excitable \cite{21a},
bistable \cite{21b} and close to bifurcations \cite{21c} systems display
oscillation behavior, whose adjacency to ideally periodic signal depends
resonantly on the noise intensity \cite{21d}, that was a reason to call these
oscillations coherence resonance \cite{21a} or stochastic coherence \cite{SSG}.
Control of the coherence resonance regime was shown to be achieved with a
time-delayed feedback, which enables to increase or decrease the regularity of
motion \cite{21e}. Characteristically, a quasioscillatory behavior may be
organized without any input signal, provided a stochastic nonlinear system has
itself an intrinsic time scale. If this scale is driven by a multiplicative
noise, which induce bistable behavior in a deterministically monostable medium,
then a doubly stochastic resonance arises \cite{Z}.

The simplest way to formulate the model related to systems with finite number
$N<\infty$ of constituents is to consider the sum ${\vec
S}=\sum_{i=1}^N{\vec\xi}_i$ of random state vectors ${\vec\xi}_i$ with
components $\xi_i^{\alpha}$, $\alpha=1,\dots,d$. Then, the state vector
\begin{equation}
 {\vec S}=N{\vec X}+\sqrt{N}{\vec x}
 \label{0}
\end{equation}
is decomposed into a deterministic component being proportional to total system
size $N$ and a random one to be proportional to its square root \cite{22}. In
the limit of infinite particle numbers $N\to\infty$, such systems are
faithfully described by deterministic equations to find time dependence ${\vec
X}(t)$, which addresses the behavior of the system on a mean-field level. On
the other hand, a systematic study of corrections due to finite system size can
capture the behavior of fluctuations ${\vec x}(t)$ about the mean-field
solution. These fluctuations are governed with the Langevin equations, however,
in difference of approach \cite{21}, we consider multiplicative noises instead
of additive ones, on the one hand, and nonlinear forces instead of linear ones,
on the other. Within such framework, the aim of the present paper is to extend
analytical descriptions \cite{21} of finite-size stochastic effects to
non-equilibrium systems where noises play a crucial role with respect to
periodic limit cycle solution creation or its supression. We will show that
character of the stationary behavior of non-equilibrium system is determined by
relation between scales of temporal variation of principle variables as well as
their coupling. In contrast to the doubly stochastic resonance \cite{Z}, we
consider the case when multistable state is caused by both multiplicative noise
and offset from equilibrium state.

The paper is organized along the following lines. In Section \ref{Sec.2}, we
obtain conditions of the limit cycle creation using pair of stochastic
equations with nonlinear forces and multiplicative noises. Sections
\ref{Sec.3}, \ref{Sec.4} are devoted to consideration of these conditions on
the basis of stochastic Lorenz system with different regimes of principle
variables slaving. According to Section \ref{Sec.3} the limit cycle is created
only in the case if the most fast variation displays a principle variable,
which is coupled nonlinearly with two other degrees of freedom or more.
Opposite case is studied in Section \ref{Sec.4} to show that the limit cycle
disappears in non-equilibrium steady state. Section \ref{Sec.5} concludes our
consideration.

\section{Statistical picture of limit cycle}\label{Sec.2}

According to the theorem of central manifold \cite{17}, to achie\-ve a closed
description of a limit cycle it is enough to use only two variables $x_\alpha$,
$\alpha=1,2$. In such a case, stochastic evolution of the system under
investigation is defined by the Langevin equations \cite{23}
\begin{equation}
 \dot{x}_\alpha=f^{(\alpha)}+\mathcal{G}_{\alpha}\zeta_{\alpha}(t),\quad \alpha=1,2
 \label{1}
\end{equation}
with forces $f^{(\alpha)}=f^{(\alpha)}(x_1,x_2)$ and noise amplitudes
$\mathcal{G}_{\alpha}=\mathcal{G}_{\alpha}(x_1,x_2)$, being functions of
stochastic variables $x_\alpha$, $\alpha=1,2$; white noises $\zeta_{\alpha}(t)$
are determined by usual conditions $\langle\zeta_{\alpha}(t)\rangle=0$,
$\langle\zeta_{\alpha}(t)\zeta_\beta(t')\rangle=\delta_{\alpha\beta}\delta(t-t')$.
Within the assumption that microscopic transfer rates are non correlated for
different variables $x_\alpha$ (see below), the probability distribution
function $\mathcal{P}=\mathcal{P}(x_1,x_2;t)$ is determined by the
Fokker-Planck equation
\begin{equation}
\frac{\partial\mathcal{P}}{\partial t}+\sum\limits_{\alpha=1}^{2}\frac{\partial
J^\alpha}{\partial x_\alpha}=0,
 \label{3}
\end{equation}
where components of the probability current take the form
\begin{equation}
J^{(\alpha)}\equiv
\mathcal{F}^{(\alpha)}\mathcal{P}-\frac{1}{2}\sum\limits_{\beta=1}^{2}\frac{\partial
}{\partial x_\beta}\left(\mathcal{G}_\alpha\mathcal{G}_\beta\mathcal{P}\right)
 \label{4}
\end{equation}
with the generalized forces
\begin{equation}
\mathcal{F}^{(\alpha)}=f^{(\alpha)}+
\lambda\sum\limits_{\beta=1}^{2}\frac{\partial\left(\mathcal{G}_\alpha\mathcal{G}_\beta
\right)}{\partial x_\beta}, \label{4a}
\end{equation}
being determined with choice of the calculus parameter $\lambda\in [0,1]$ (for
Ito and Stratonovich cases, one has $\lambda=0$ and $\lambda=1/2$,
respectively). Within the steady state, the components of the probability
current take constant values $J^{(\alpha)}_0$ and the system behaviour is
defined by the following equations:
\begin{eqnarray} \label{6}
 \frac{\partial}{\partial
x_1}\left(\mathcal{G}_1^2\mathcal{P}\right)+\frac{\partial }{\partial
x_2}\left(\mathcal{G}_1\mathcal{G}_2\mathcal{P}\right)-2\mathcal{F}^{(1)}\mathcal{P}=-2J^{(1)}_0,\\
 \frac{\partial }{\partial
x_1}\left(\mathcal{G}_1\mathcal{G}_2\mathcal{P}\right)+\frac{\partial
}{\partial
x_2}\left(\mathcal{G}_2^2\mathcal{P}\right)-2\mathcal{F}^{(2)}\mathcal{P}=-2J^{(2)}_0.
\end{eqnarray}

Multiplying the first of these equations by factor $\mathcal{G}_2$ and the
second one by $\mathcal{G}_1$ and then subtracting results, we arrive at the
explicit form of the probability distribution function as follows:
\begin{eqnarray} \label{7}
\mathcal{P}\left(x_1,x_2\right)=\frac{J^{(1)}_0\mathcal{G}_2\left(x_1,x_2\right)
-J^{(2)}_0\mathcal{G}_1\left(x_1,x_2\right)}{\mathcal{D}\left(x_1,x_2\right)},\nonumber\\
\mathcal{D}\left(x_1,x_2\right)\equiv\left(\mathcal{G}_2\mathcal{F}^{(1)}
-\mathcal{G}_1\mathcal{F}^{(2)}\right)\\
+\frac{1}{2}\left[\left(\mathcal{G}_1^2\frac{\partial\mathcal{G}_2}{\partial
x_1}-\mathcal{G}_2^2\frac{\partial\mathcal{G}_1}{\partial
x_2}\right)-\mathcal{G}_1\mathcal{G}_2\left(\frac{\partial\mathcal{G}_1}{\partial
x_1}-\frac{\partial\mathcal{G}_2}{\partial x_2}\right)\right].\nonumber
\end{eqnarray}
This function diverges at condition
\begin{equation}
2\left(\mathcal{G}_1\mathcal{F}^{(2)}-\mathcal{G}_2\mathcal{F}^{(1)}\right)\nonumber
=\left(\mathcal{G}_1^2\frac{\partial\mathcal{G}_2}{\partial
x_1}-\mathcal{G}_2^2\frac{\partial\mathcal{G}_1}{\partial
x_2}\right)-\mathcal{G}_1\mathcal{G}_2\left(\frac{\partial\mathcal{G}_1}{\partial
x_1}-\frac{\partial\mathcal{G}_2}{\partial x_2}\right),
 \label{8}
\end{equation}
that physically means appearance of a domain of forbidden values of stochastic
variables $x_\alpha$, which is bonded with a closed line of the limit cycle.
Characteristically, such a line appears only if the denominator
$\mathcal{D}(x_1,x_2)$ of fraction (\ref{7}) includes even powers of both
variables $x_1$ and $x_2$.\footnote{Archetype of closed curves presents the
circle $x_1^2+x_2^2=1$.}

It is worth to note that the analytical expression (\ref{7}) of the probability
distribution function becomes possible due to the special form of the
probability current (\ref{4}), where effective diffusion coefficient takes the
multiplicative form
$\mathcal{D}_{\alpha\beta}=\mathcal{G}_\alpha\mathcal{G}_\beta$. In general
case, this coefficient is known to be defined with the expression \cite{24}
\begin{equation}
\mathcal{D}_{\alpha\beta}=\sum\limits_{ab}I_{ab}g_\alpha^ag_\beta^b,
 \label{8a}
\end{equation}
where kernel $I_{ab}$ determines transfer rate between microscopic states $a$
and $b$, whereas factors $g_\alpha^a$ and $g_\beta^b$ are specific noise
amplitudes of values $x_\alpha$ related to these states. We have considered
above the simplest case, when the transfer rate $I_{ab}=I$ is constant for all
microscopic states. As a result, the diffusion coefficient (\ref{8a}) takes the
needed form $\mathcal{D}_{\alpha\beta}=\mathcal{G}_\alpha\mathcal{G}_\beta$
with cumulative noise amplitudes $\mathcal{G}_\alpha
\equiv\sqrt{I}\sum_{a}g_\alpha^a$ and
$\mathcal{G}_\beta\equiv\sqrt{I}\sum_{b}g_\beta^b$.

\section{Noise induced resonance within Lorenz system}\label{Sec.3}

As the simplest and most popular example of the self-organization induced by
the Hopf bifurcation, we consider modulation regime of spontaneous laser
radiation, whose behaviour is presented in terms of the radiation strength $E$,
the matter polarization $P$ and the difference of level populations $S$
\cite{22}. With accounting for stochastic sources related, the
self-organization process of this system is described by the Lorenz equations
\begin{eqnarray}\label{lor}
 \tau_E\dot{E}=[-E+a_E P-\varphi(E)]+g_E\zeta(t),\nonumber\\
 \tau_P\dot{P}=(-P+a_P ES)+g_P\zeta(t),\\
 \tau_S\dot{S}=[(S_e-S)-a_S EP]+g_S\zeta(t).\nonumber
\end{eqnarray}
Here, overdot denotes differentiation over time $t$; $\tau_{E,P,S}$ and
$a_{E,P,S}>0$ are time scales and feedback constants of related variables,
respectively; $g_{E,P,S}$ are corresponding noise amplitudes, and $S_e$ is
driven force. In the absence of noises $(g_{E}=g_{P}=g_{S}=0)$ and at relations
$\tau_P,\tau_S\ll\tau_E$ between time scales, the system (\ref{lor}) addresses
to limit cycle only in the presence of the nonlinear force \cite{25}
\begin{equation}
\varphi(E)=\frac{\kappa E}{1+E^2/E_n^2} \label{11ab}
\end{equation}
characterized with parameters $\kappa>0$ and $E_n$. In this Section, we
consider noise effect in the case of opposite relations
$\tau_E\ll\tau_P,\tau_S$ of time scales, when periodic variation of stochastic
variables becomes possible even at suppression of the force (\ref{11ab}).

It is convenient further to pass to dimensionless variables $t$, $\zeta$, $E$,
$P$, $S$, $g_E$, $g_P$, $g_S$ with making use of the related scales:
\begin{eqnarray} \label{scale}
\tau_P;\ \zeta_s=\tau_P^{-1/2};\ E_{s}=(a_Pa_S)^{-1/2},\ P_s=(a^2_E a_P
a_S)^{-1/2},\ S_s=(a_E a_P)^{-1};\nonumber\\ g_E^{s}=(\tau_P/a_Pa_S)^{1/2},\
g_P^s=(\tau_P/a^2_E a_P a_S)^{1/2},\ g_S^s=\tau_P^{1/2}/a_E a_P.
\end{eqnarray}
Then, equations (\ref{lor}) take the simple form\footnote{These equations are
reduced to the initial Lorenz form \cite{22a} if we set
$X\equiv\sqrt{\sigma/\varepsilon}E$, $Y\equiv\sqrt{\sigma/\varepsilon}P$,
$Z\equiv S_e-S$, $r\equiv S_e$, $b\equiv\sigma/\varepsilon$ and
$g_E=g_P=g_S=0$.}
\begin{eqnarray}\label{lor11a}
 \sigma^{-1}\dot{E}=-E+P-\varphi(E)+g_E\zeta(t),\nonumber\\
 \dot{P}=-P+ES+g_P\zeta(t),\\
 (\varepsilon/\sigma)\dot{S}=(S_e-S)-EP+g_S\zeta(t)\nonumber,
\end{eqnarray}
where the time scale ratios
\begin{equation}
\sigma=\tau_P/\tau_E,\quad
 \varepsilon=\tau_S/\tau_E
 \label{time}
\end{equation}
are introduced. In the absence of the noises, the Lorenz system (\ref{lor11a})
is known to show the usual bifurcation in the point $S_e=1$ and the Hopf
bifurcation at the driven force \cite{22a,22}
\begin{equation}
S_e=\frac{\tau_P}{\tau_E}~\frac{\tau_E^{-1}+\tau_S^{-1}+3\tau_P^{-1}}
{\tau_E^{-1}-\tau_S^{-1}-\tau_P^{-1}}.
 \label{cond}
\end{equation}
However, the noiseless limit cycle $(g_E=g_P=g_S=0)$ is unstable and the Hopf
bifurcation arrives at the strange attractor only.

With switching on the noises, the condition $\tau_E\ll\tau_P$ allows for to put
l.h.s. of the first equation (\ref{lor11a}) to be equal zero. Then, the
radiation strength is expressed with the equality
\begin{equation}
E=P+g_E\zeta(t), (\ref{11a})
 \label{11a}
\end{equation}
whose insertion into the system (\ref{lor11a}) reduces it into two-dimensional
form
\begin{eqnarray}\label{22}
 \dot{P}=-P(1-S)+\mathcal{G}_P\zeta(t),\nonumber\\
 \dot{S}=(\sigma/\varepsilon)\left[(S_e-S)-P^2\right]+\mathcal{G}_S\zeta(t)
 \end{eqnarray}
with the effective amplitudes of multiplicative noises
\begin{equation}\label{11abc}
\mathcal{G}_P=\sqrt{g_P^2+g_E^2S^2},\quad
\mathcal{G}_S=(\tau_{P}/\tau_{S})\sqrt{g_S^2+g_E^2P^2}
\end{equation}
and the generalized forces
\begin{eqnarray}
 \mathcal{F}^{(P)}=-P(1-S)+\lambda\frac{g_E^2}{\tau_{S}/\tau_{P}}S
 \sqrt{\frac{(g_S/g_E)^2+P^2}{(g_P/g_E)^2+S^2}},\nonumber\\
 \mathcal{F}^{(S)}=(\tau_{P}/\tau_{S})\left[(S_e-S)-P^2\right]
 +\lambda\frac{g_E^2}{\tau_{S}/\tau_{P}}P
 \sqrt{\frac{(g_P/g_E)^2+S^2}{(g_S/g_E)^2+P^2}}.
\end{eqnarray}
In this way, the probability density (\ref{7}) takes infinite values at
condition
\begin{eqnarray} \label{111}
\left(\frac{g_S^2}{g_E^2}+P^2\right)\sqrt{\frac{g_P^2}{g_E^2}+S^2}P(1-S)+
\left(\frac{g_P^2}{g_E^2}+S^2\right)\sqrt{\frac{g_S^2}{g_E^2}+P^2}\left[(S_e-S)-P^2\right]\nonumber\\+
\frac{g_E^2}{2}\frac{\sigma}{\varepsilon}\left(\frac{g_S^2}{g_E^2}
+P^2\right)^{\frac{3}{2}}S-\frac{g_E^2}{2}
\left(\frac{g_P^2}{g_E^2}+S^2\right)^{\frac{3}{2}}P=0,
\end{eqnarray}
where we choose the simplest case of the Ito calculus $(\lambda=0)$.

\begin{figure}
\centering
\includegraphics[width=0.48\textwidth]{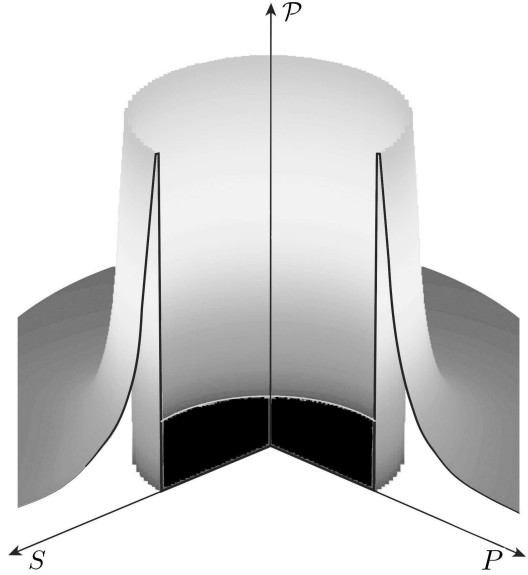}
\caption{Steady state distribution function (\ref{7}) at
$J^{(P)}_0=1$, $J^{(S)}_0=10$, $\tau_P=\tau_S$, $S_e=0.5$,
$g_E=0.5$, $g_P=1.376$, $g_S=2.5$}\label{prob1.eps}
\end{figure}

\begin{figure}
\centering a\hspace{0.5\textwidth}b\\
\includegraphics[width=0.45\textwidth]{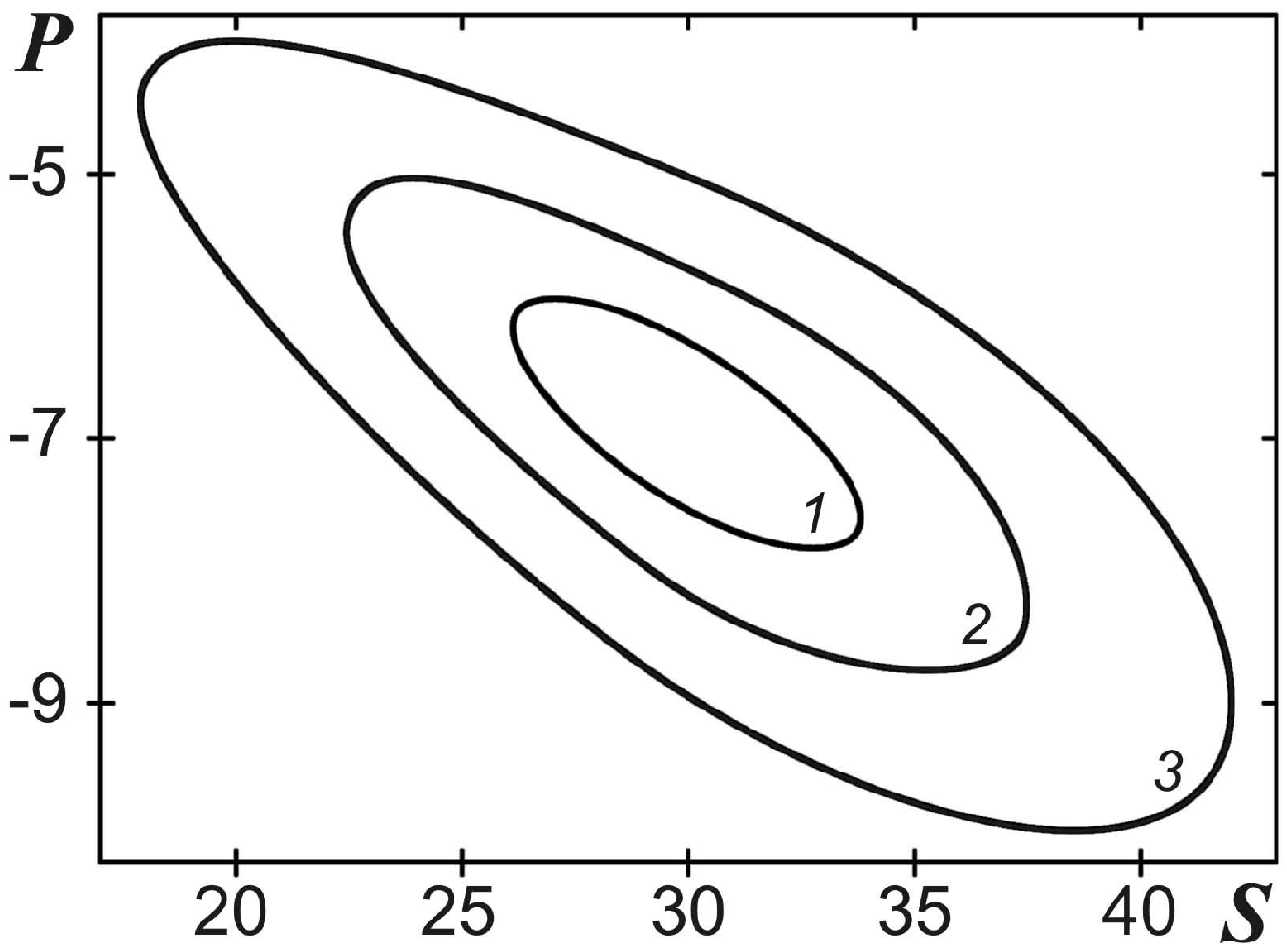}
\includegraphics[width=0.45\textwidth]{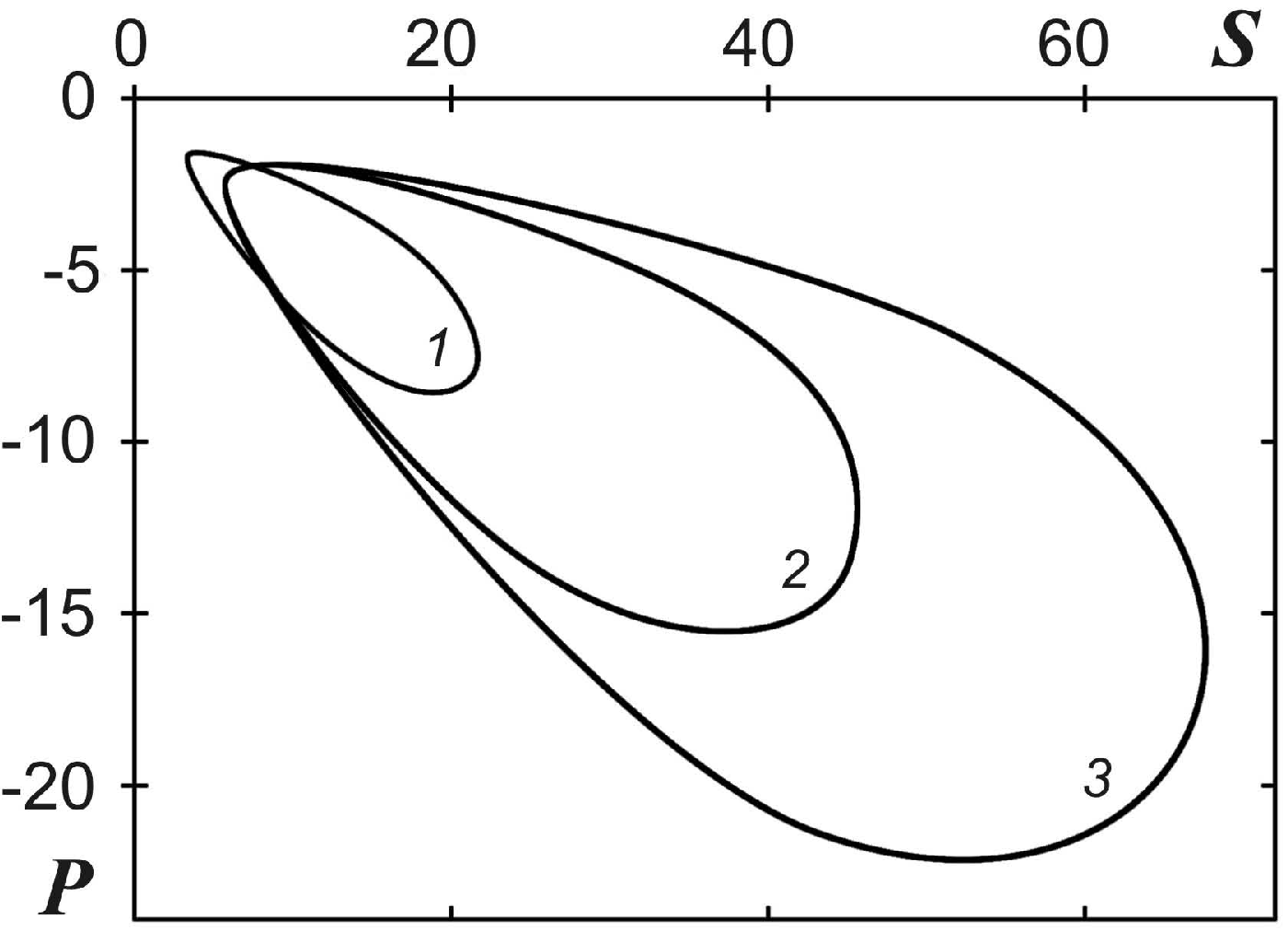}
\caption{Form of limit cycle determined with Eq.(\ref{111}) at
$\varepsilon=1$, $\sigma=1$ and: \break a) $g_E=0.5$, $g_P=11$,
$g_S=6$ (curves 1-3 relate to $S_e=0.5, 1.0, 2.0$, respectively);
\break b) $S_e=0.5$, $g_P=7.5$, $g_S=6.5$ (curves 1-3 relate to
$g_E=1.0, 0.6, 0.5$, respectively)}\label{ab.eps}
\end{figure}

\begin{figure}
\centering a\hspace{0.5\textwidth}b\\
\includegraphics[width=0.45\textwidth]{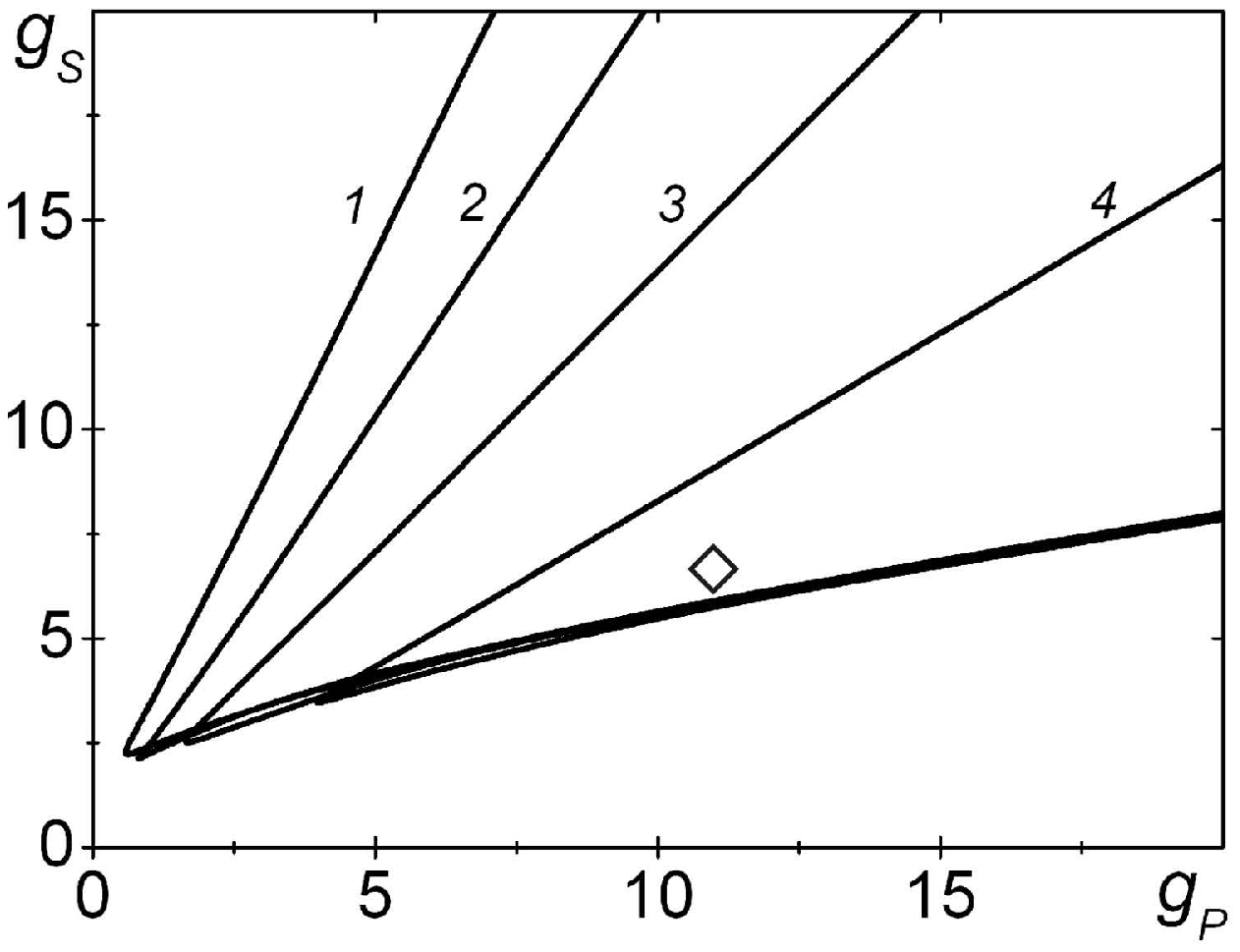}
\includegraphics[width=0.45\textwidth]{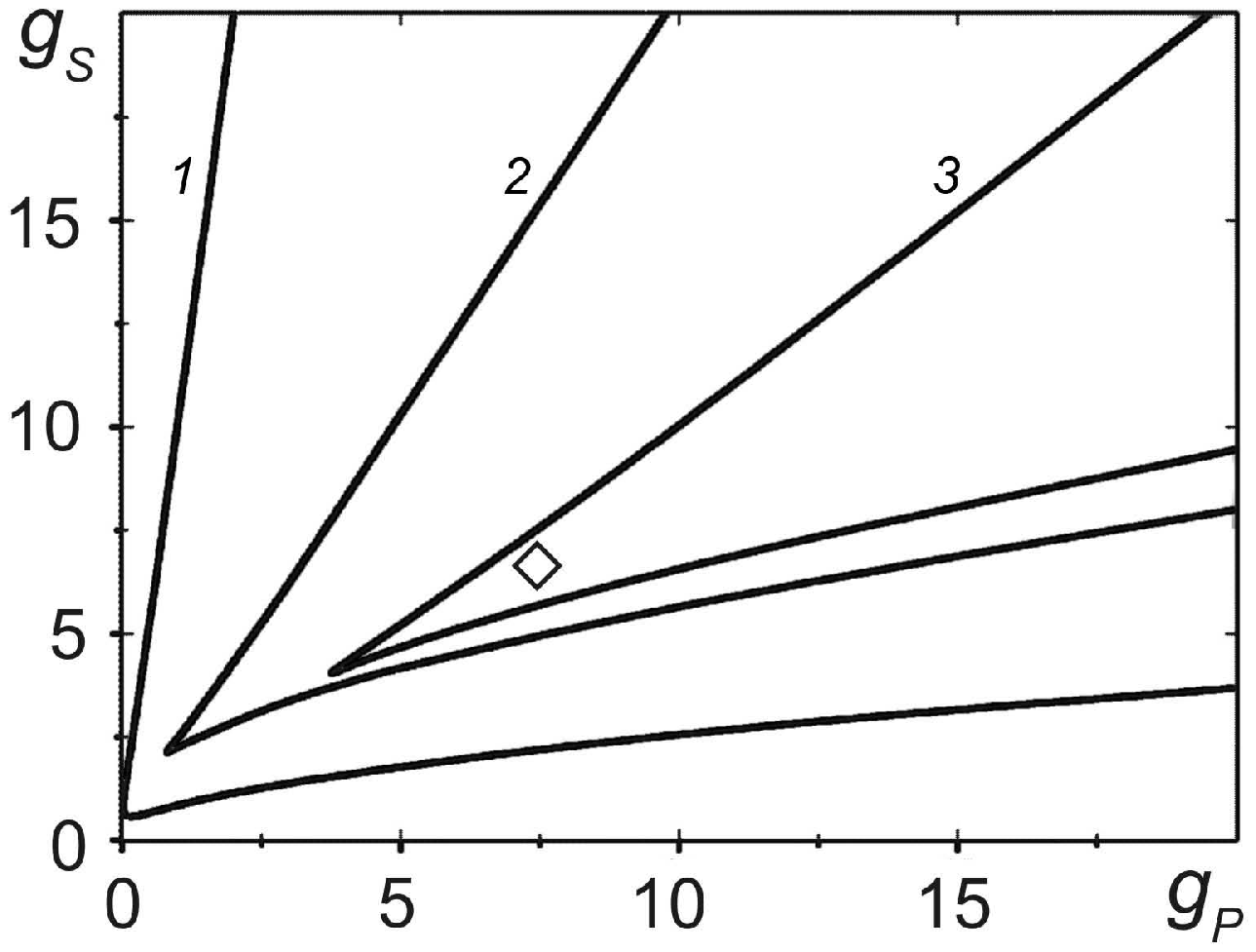}
\caption{Phase diagrams of the limit cycle creation at
$\varepsilon=1$, $\sigma=1$ and: a) $g_E=0.5$, curves 1-4
correspond to $S_e=0.0, 0.5, 1.0, 2.0$, respectively; b)
$S_e=0.5$, curves 1-3 correspond to $g_E=0.1, 0.5, 1.0$,
respectively (diamonds relate to the values $g_P$ and $g_S$, for
which limit cycles in Fig.\ref{ab.eps} are
depicted)}\label{se.eps}
\end{figure}

Reduced Lorenz system (\ref{22}) has two-dimensional form
(\ref{1}), where the role of variables $x_1$ and $x_2$ play the
matter polarization $P$ and the difference of level populations
$S$. According to the distribution function (\ref{7}) shown in
Fig.\ref{prob1.eps}, the stochastic
variables $P$ and $S$ are realized with non-zero probabilities out
off the limit cycle only, whereas in its interior the domain of
forbidden values $P$, $S$ appears. That is principle difference
from the deterministic limit cycle, which bounds a domain of
unstable values of related variables. The form of this domain is
shown in Fig.\ref{ab.eps}
at different values of the noise amplitudes $g_E$, $g_P$, $g_S$
and driven force $S_e$. It is seen, this domain grows with
increase of the driven force $S_e$, whereas increase of the force
fluctuations $g_E$ shrinks it. On the other hand, phase diagrams
depicted in Fig.\ref{se.eps} show
that increasing the noise amplitudes of both polarization and difference of
level populations enlarges domain of the limit cycle creation (more exactly,
the noise amplitude $g_E$ shrinks this domain from both above and below,
whereas increase of the driven force $S_e$ makes the same from above only).

The principle peculiarity of the limit cycles obtained is that their form,
determined with Eq.(\ref{111}), does not depend on a non-equilibrium degree
fixed by the stationary probability currents $J_0^{(1,2)}$, whereas the
probability (\ref{7}) itself does not equal zero at conditions $J_0^{(1,2)}\ne
0$ only. In this connection, one should be pointed out a non-triviality of the
problem of numerical solution of the reduced Lorenz system (\ref{22}), which
determines these limit cycles initially. Indeed, resolving this problem
proposes the following steps: i) direct solution of the stochastic equations
(\ref{22}) to find a set of the time dependencies $P(t)$ and $S(t)$; ii)
numerical determination of the time-dependent probability $\mathcal{P}(P,S;t)$
to realize entire set of possible solutions of the equations (\ref{22}); iii)
selection of non-equilibrium solutions, which obey to the steady state
condition $J^{(\alpha)}=J_0^{(\alpha)}$, $\alpha=1,2$, determined with the
probabillity current (\ref{4}); iv) calculation of the probability distribution
$\mathcal{P}_s(P,S)$ of the steady state solutions; v) determination of the
stochastic limit cycle according to the condition $\mathcal{P}_s(P,S)=\infty$.
Realization of this program is in progress.

\section{Lorenz system without limit cycle}\label{Sec.4}

According to Ref. \cite{25}, at conditions $\tau_{P}\ll\tau_{E},\tau_{S}$, the
deterministic system $(g_{E,P,S}=0)$ has a limit cycle only at large intensity
$\kappa$ of non linear force (\ref{11ab}). In this case, it is convenient to
measure the time $t$ in the scale $\tau_E$ and replace $\tau_P$ by $\tau_E$ in
set of scales (\ref{scale}). Then, one obtains instead of Eq.(\ref{11a}) the
relation
\begin{equation}
P=ES+g_P\zeta(t),
\end{equation}
due to which the Lorenz system (\ref{lor11a}) is reduced to two-dimensional
form
\begin{eqnarray}
 \dot{ E}=-\left[E(1-S)+\varphi(E)\right]+\mathcal{G}_E\zeta(t),\nonumber\\
 \dot{S}=\varepsilon^{-1}\left[S_e-S(1+ E^2)\right]+\mathcal{G}_S\zeta(t)\label{2b}
\end{eqnarray}
with the effective noise amplitudes
\begin{eqnarray}
\mathcal{G}_E=\sqrt{g_P^2+g_E^2},\quad
\mathcal{G}_S=\varepsilon^{-1}\sqrt{g_S^2+g_P^2 E^2}.
 \label{3b}
\end{eqnarray}
The generalized forces are as follows:
\begin{eqnarray}\label{5b}
 \mathcal{F}^{(E)}=-\left[E(1-S)+\varphi(E)\right],\nonumber\\
 \mathcal{F}^{(S)}=\varepsilon^{-1}\left[(S_e-S)-SE^2\right]
 +\lambda\frac{g_P^2}{\varepsilon}E
 \sqrt{\frac{1+(g_E/g_P)^2}{(g_S/g_P)^2+E^2}}.
\end{eqnarray}
The probability distribution function (\ref{7}) diverges at condition
\begin{eqnarray} \label{A}
\frac{\left(g_S/g_P\right)^2+E^2}{1+\left(g_E/g_P\right)^2}\left[\varphi(E)+E(1-S)\right]\nonumber\\+
\sqrt{\frac{\left(g_S/g_P\right)^2+E^2}{1+\left(g_E/g_P\right)^2}}\left[S_e-S(1+E^2)\right]
\left(\lambda-\frac{1}{2}\right)g_P^2E=0,
\end{eqnarray}
being the equation, which does not include even powers of the variable $S$.

\begin{figure}
\includegraphics[width=0.9\textwidth]{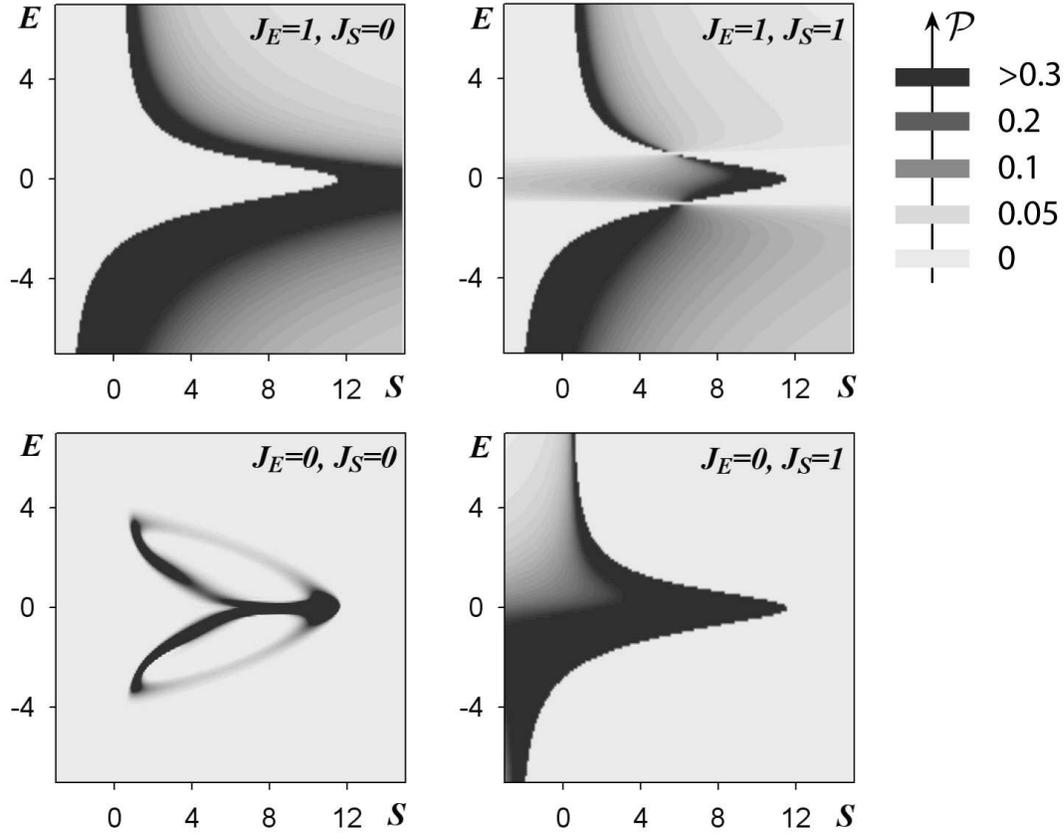}
\caption{Steady state probability distribution function in
dependence of the radiation strength $E$ and the difference of
level populations $S$ at conditions
$\tau_{P}\ll\tau_{E}=\tau_{S}$, $\kappa=10$, $S_e=11.6$,
$g_E=0.2$, $g_P=0.2$, $g_S=0.2$ and different probability currents
$J_0^{(E)}$, $J_0^{(S)}$ (shown on the panels
related)}\label{prob.eps}
\end{figure}

As a result, one can conclude that offset from equilibrium steady
state destroys a deterministic limit cycle at the relations
$\tau_{P}\ll\tau_{E},\tau_{S}$ between characteristic scales. This
conclusion is confirmed with Fig.\ref{prob.eps}, which shows
divergence of the
probability distribution function on the limit cycle of variation of the
radiation strength $E$ and the difference of level populations $S$ at zeros
probability currents $J_0^{(E)}$ and $J_0^{(S)}$ only. With increase of these
currents the system escapes from equilibrium steady state and maximum of the
distribution function shifts to non-closed curves to be determined with
equation (\ref{A}).

\section{Conclusion}\label{Sec.5}

We have considered effect of stochastic sources upon self-organization process
being initiated with creation of the limit cycle. In Sections \ref{Sec.3},
\ref{Sec.4}, we have applied general relations obtained in Section \ref{Sec.2}
to the stochastic Lorenz system. We have shown that offset from equilibrium
steady state can destroy or create the limit cycle in dependence of relation
between characteristic scales of temporal variation of principle variables.

Investigation of the Lorenz system with different re\-gi\-mes of principle
variables slaving shows that additive noises can take multiplicative character
if one of these noises has many fewer time scale than others. In such a case,
the limit cycle may be created if the most fast variable is coupled with more
than two slow ones. However, the case considered in Section \ref{Sec.4} shows
that such dependence is not necessary to arrive at limit cycle. The formal
reason is that within adiabatic condition $\tau_P\ll\tau_E$ both noise
amplitude $\mathcal{G}_S(E)$ and generalized force $\mathcal{F}^{(S)}(E)$,
determined with Eqs. (\ref{3b}) and (\ref{5b}), enclose the squared strength
$E^2$, but do not include the square $S^2$.

The limit cycle is created if the fastest variations displays a principle
variable, which is coupled with two different degrees of freedom or more.
Indeed, at the relations $\tau_E\ll\tau_P,\tau_S$ of relaxation times
considered in Section \ref{Sec.3}, the strength $E$ evolves according to the
stochastic law of motion (\ref{11a}). Accounting this relation in the nonlinear
terms of two last equations (\ref{lor11a}) arrives at dependencies of the noise
amplitudes of the polarization $P$ and the difference of level populations $S$
on both variables $S$ and $P$ themselves. Due to gaussian nature of the noises
their variances are additive values \cite{23,24}, so that effective noise
amplitudes $\mathcal{G}_P$ and $\mathcal{G}_S$ of the principle variables are
defined by Eqs. (\ref{11abc}), which include both squares $S^2$ and $P^2$. As a
result, solutions of Eq.(\ref{111}) become double-valued to be related to the
limit cycle.

This cycle appears physically as stochastic coherence, that has been observed
both numerically \cite{d,21a} and analytically \cite{26}. Analogously to the
doubly stochastic resonance \cite{Z}, such a resonance may be organized if
stochastic nonlinear system has two noises, but both of them must be
multiplicative in nature. Moreover, the system under study acquires an
intrinsic time scale related to multistable state only far off equilibrium
statistical state. In opposite to the deterministic limit cycle which bounds a
domain of unstable values of related variables, in our case stochastic
variables evolve out off the limit cycle only, whereas in its interior the
domain of forbidden values appears.

\section{Acknowledgements}\label{Sec.6}

We are grateful to anonymous referee for constructive criticism and bringing to
our attention Refs. \cite{21a,21d,21e}. We thank also Prof. Jos\'e M. Sancho
for Ref. \cite{SSG}.

\end{document}